\documentclass[prl,twocolumn,preprintnumbers,longbibliography,amsmath,amssymb,nobibnotes,superscriptaddress]{revtex4}

\usepackage{graphicx}% Include figure files
\graphicspath{{Figures/PNG files}{./}}
\usepackage{dcolumn}% Align table columns on decimal point
\usepackage{bm}% bold math
\usepackage{float}
\usepackage{hyphenat}
\usepackage[dvipsnames]{xcolor}

\newcommand{\units}[1]% = \unit sans l'espace devant
{\text{#1}}

\newcommand{\diff}{\mathrm{d}}

\newcommand{\valencia}{ICMUV, Instituto de Ciencia de Materiales, Universidad de Valencia, P.O. Box 22085, 46071 Valencia, Spain}
\newcommand{\heriotwatt}{Institute of Photonics and Quantum Sciences, SUPA, Heriot-Watt University, Edinburgh EH14 4AS, UK}
\newcommand{\stockholm}{Current address: Department of Applied Physics, Royal Institute of Technology, Stockholm 106 91, Sweden}
\newcommand{\equalcontrib}{These authors contributed equally to this work}
\newcommand{\inl}{International Iberian Nanotechnology Laboratory (INL), Av. Mestre Jos\'e Veiga, 4715-330 Braga, Portugal}

\begin{document}

	\title{Out-of-plane orientation of luminescent excitons in two-dimensional indium selenide}
	
	\author{Mauro Brotons-Gisbert}
	\thanks{\equalcontrib}
	\email{M.Brotons_i_Gisbert@hw.ac.uk}
	\affiliation{\heriotwatt}
	\author{Rapha\"{e}l Proux}
	\thanks{\equalcontrib}
	\email{M.Brotons_i_Gisbert@hw.ac.uk}
	\affiliation{\heriotwatt}
	\author{Rapha\"{e}l Picard}
	\affiliation{\heriotwatt}
	\author{Daniel Andres-Penares}
	\affiliation{\valencia}
	\author{Artur Branny}
	\affiliation{\heriotwatt}
	\affiliation{\stockholm}
	\author{Alejandro Molina-S\'{a}nchez}
	\affiliation{\valencia}
	\affiliation{\inl}
	\author{Juan F. S\'{a}nchez-Royo}
	\email{Juan.F.Sanchez@uv.es}
	\affiliation{\valencia}
	\author{Brian D. Gerardot}
	\email{B.D.Gerardot@hw.ac.uk}
	\affiliation{\heriotwatt}

	\begin{abstract}
		\section{Abstract}
		Van der Waals materials offer a wide range of atomic layers with unique properties that can be easily combined to engineer novel electronic and photonic devices. A missing ingredient of the van der Waals platform is a two-dimensional crystal with naturally occurring out-of-plane luminescent dipole orientation. Here we measure the far-field photoluminescence intensity distribution of bulk InSe and two-dimensional InSe, WSe$_2$ and MoSe$_2$. We demonstrate, with the support of ab-initio calculations, that layered InSe flakes sustain luminescent excitons with an intrinsic out-of-plane orientation, in contrast with the in-plane orientation of dipoles we find in two-dimensional WSe$_2$ and MoSe$_2$ at room-temperature. These results, combined with the high tunability of the optical response and outstanding transport properties, position layered InSe as a promising semiconductor for novel optoelectronic devices, in particular for hybrid integrated photonic chips which exploit the out-of-plane dipole orientation.
	\end{abstract}

	\maketitle
	\section{Introduction}
	The creation and recombination of excitons, electron-hole pairs bound by Coulomb forces \cite{Frenkel1931a}, mediates light-matter interaction in semiconductors. The exciton transition-dipole moment can be highly anisotropic, with the dipole's strength and orientation dictated by the particular electronic properties of the host semiconductor and its dipole selection rules \cite{Yu2005Fundamentals, chen2018theory}. This fundamental relationship implies that different low-dimensional semiconductor structures with varying electronic properties can be found to yield in-plane (IP), out-of-plane (OP), or mixed dipole orientations. Further, beyond purely electronic effects, an anisotropic crystal shape also induces additional dielectric effects which affect the intrinsic transition dipole moment \cite{Scott2017}. 
	
	The orientation of radiative excitons determines their potential in optoelectronic and integrated photonic applications. IP dipoles are more easily accessed in experiments, but emerging applications in integrated photonic chips can be enabled with OP orientations to efficiently mediate light-matter interactions with planar and cylindrical waveguides \cite{wang2016surface, jun2009broadband, rong2014spontaneous, verhart2014single, davancco2009efficient, liebermeister2014tapered}. Recently, two-dimensional (2D) transition metal dichalcogenide (TMD) semiconductors which sustain robust excitons have emerged \cite{molina2013effect, chernikov2014exciton}, presenting new opportunities to engineer light-matter interaction at the nanoscale. Typically, these 2D semiconductors present bright luminescent intra-layer excitons with IP transitions which favour directional out-coupling of radiation, as has been demonstrated in MoS$_2$ at room temperature \cite{Schuller2013}. In TMDs, OP transitions can only be found in weakly luminsecent (so-called grey) excitons \cite{wang2017plane, zhou2017probing, zhang2017magnetic} or engineered structures such as chemically transformed Janus monolayers (MLs) \cite{Lu2017,Zhang2017} or inter-layer excitons in type II heterostructures which have small oscillator strength \cite{hong2014ultrafast, fang2014strong,Rivera2016,torun2018interlayer}.
	
	In contrast to TMDs, layered III-VI chalcogenides have a band-gap defined by intra-layer electronic states with a major $p_z$ atomic orbital composition. This endows InSe \cite{camassel1978excitonic, mudd2013tuning, sanchez2014electronic, brotons2016nanotexturing}, GaSe \cite{andres2017quantum, terry2018infrared} and their heterostructures \cite{terry2018infrared} with large quantum confinement effects: as the layer thickness varies from the bulk to ML, the band-gaps tune from the infrared to violet. The large band-gap tunability can be exploited for optoelectronic applications \cite{tamalampudi2014high, feng2015ultrahigh, mudd2015high} while the broken inversion symmetry of InSe crystals leads to strong optical nonlinearities even at the atomic thickness limit \cite{leisgang2018optical}. Narrow quantum dot-like emissions that have been observed in both GaSe \cite{tonndorf2017single, tonndorf2017chip} and InSe \cite{mudd2016direct} offer promise for future quantum photonic devices. Electrons in the conduction band of InSe have small effective mass \cite{kuroda1980resonance, kress1982cyclotron} and weak electron-phonon scattering \cite{segura1984electron} which combine to yield high electron mobility, in particular in devices encapsulated by hexagonal boron nitride \cite{bandurin2017high}. These features make InSe uniquely attractive among the wide array of layered van der Waals materials. Additionally, the $p_z$ atomic orbital nature, combined with the symmetry of the valence and conduction band states at the band-gap of layered InSe, may naturally provide for excitons with large oscillator strengths and large OP dipole orientation \cite{camassel1978excitonic, mudd2013tuning, sanchez2014electronic, brotons2016nanotexturing, Magorrian2016, bandurin2017high}. 
	
	Recent experimental works have attempted to demonstrate the OP dipole orientation of InSe \cite{brotons2016nanotexturing, li2018enhanced}. Unfortunately, those experiments do not allow unambiguous observation of the OP dipole orientation. The strong effects of strain \cite{song2018largely} and dielectric-induced luminescence enhancement \cite{lien2015engineering, brotons2018engineering} arise, which can be competing explanations. More importantly, the concept behind these works does not allow a quantitative determination of the exact dipole orientation. Here we report the far-field photoluminescence (PL) intensity distribution from ML WSe$_2$ and MoSe$_2$, representative of the TMD semiconductor family, and InSe samples of varying thicknesses down to the quantum confinement regime. We find that the room-temperature PL from WSe$_2$ and MoSe$_2$ originates exclusively from IP dipoles, which indicates that grey excitons in these semiconductors (with OP dipoles) do not contribute significantly to the room-temperature emission. This result contrasts with the reported low-temperature PL of W-based TMDs such as WSe$_2$ and WS$_2$, in which OP dipoles contribute significantly to the emitted PL signal \cite{wang2017plane}. In contrast to 2D TMDs, our experimental results reveal that layered InSe flakes sustain luminescent excitons with an intrinsic OP orientation.We perform \textit{ab-initio} calculations of the electronic band structure of bulk and 2D InSe to reveal the dipole orientation is due to the behavior under the inversion symmetry operation of the electronic states, unique to the III-VI semiconductor family. Our results represent an unambiguous quantification of the orientation of the emission dipole of InSe flakes with different thicknesses. The experimental technique used in this work allows us to disentangle the dipole orientation from several additional effects that have prevented a reliable and quantitative determination of the dipole orientation of InSe to date.
	
	\section{RESULTS}
	\subsection{Orientation of luminescent excitons in bulk InSe}
	Figure 1a summarises the experimental concept used to determine the three-dimensional orientation of luminescent excitons in the van der Waals materials we investigate. The far-field PL emission pattern at room temperature is recorded by collecting the emitted photons with a microscope objective and imaging the intensity distribution in the objective's back-focal plane on a CCD camera. The inclusion of a linear polariser between the objective and the CCD allows separation of \textit{s}- and \textit{p}-polarised wave components to the back-focal plane (\textbf{k}-space). The numerical aperture (NA) of the objective was 0.95. A more detailed scheme of the experimental setup can be found in Supplementary Figure 1. This experimental technique has been extensively used to determine the radiation patterns of different fluorescent systems such as single molecules \cite{lieb2004single} and single photon emitters \cite{curto2010unidirectional,lee2011planar, bulgarini2014nanowire}. More recently, back-focal-plane imaging has also been employed to resolve the orientation of luminescent excitons in 2D semiconductors such as MoS$_2$ \cite{Schuller2013} and CdSe nanoplatelets \cite{Scott2017, gao2017cdse, heckmann2017directed}.
	
	\begin{figure*}[t]
		\begin{center}
		\includegraphics[scale=0.6]{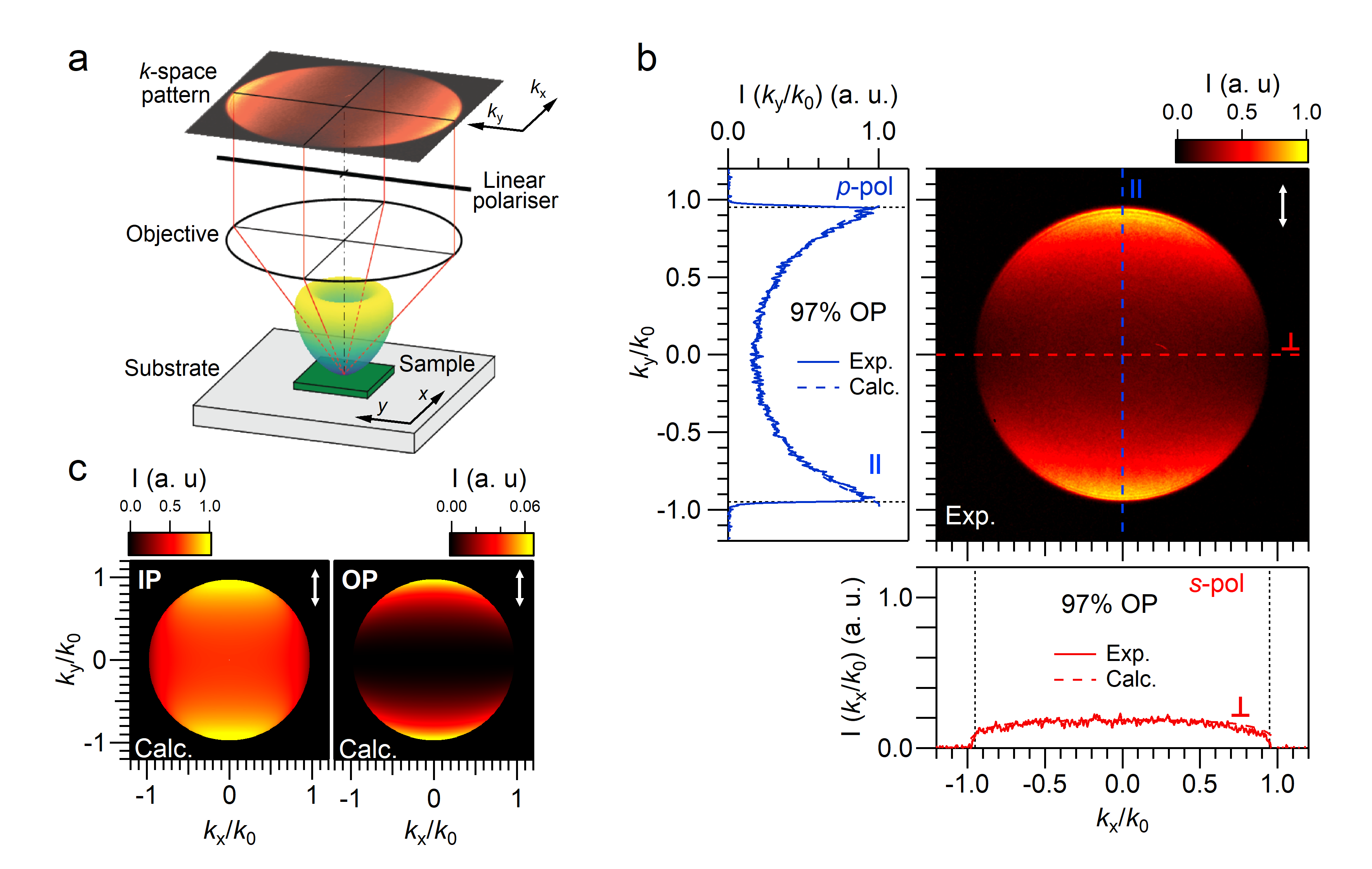}
		\end{center}
		\caption{\textbf{Orientation of luminescent excitons in bulk InSe.}  \textbf{(a)} Sketch summarising  the basic concept of the \textbf{k}-space spectroscopy. \textbf{(b)} Intensity-normalised \textbf{k}-space emission pattern of a 90 nm-thick InSe flake deposited on top of a SiO$_2$/Si substrate with a SiO$_2$ thickness of 105.3 $\pm$ 0.1 nm as a function of the in-plane photon wavevector normalised to that in air. The orientation of the transmission axis of the linear polariser used during the experiment is indicated by the white arrow in the top right corner. Blue and red solid lines in the left and bottom panel show the experimental \textbf{k}-space cross-sections measured along the directions $\parallel$ and $\perp$ to the polariser (blue and red dashed lines in the CCD image, respectively). The blue and red dashed lines in the adjacent panels represent fittings of the experimental data to the analytical model described in Methods, while the vertical black dashed lines represent the largest radiation wavevector collected by the NA of the objective. \textbf{(c)} Normalised \textbf{k}-space emission patterns calculated for pure IP (left panel) and pure OP (right panel) distributions of dipoles, respectively, emitting at an energy of $\sim$1.244 eV (see Fig. 2a) and distributed all along the thickness of a 90 nm-thick InSe flake deposited on top of a SiO$_2$/Si substrate with a SiO$_2$ thickness of 105.3 nm.}
		\label{figure1}
	\end{figure*}
	
	Figure 1b  shows the normalised \textbf{k}-space emission pattern of a mechanically exfoliated 90-nm-thick InSe bulk flake deposited on top of a SiO$_2$/Si substrate with a SiO$_2$ thickness of 105.3 
	$\pm$ 0.1 nm, as measured by nulling ellipsometry (see Methods). The vertical ($k_{y}/k_{0}$) and horizontal ($k_{x}/k_{0}$) axes represent the orthogonal components of the in-plane photon wavevector ($k_0$sin($\theta$), with $\theta$ being the emission angle) normalised to the photon wavevector in air ($k_0$). Radiation with an in-plane wavevector larger than the NA of the objective  is not collected by the objective (black region). The white arrow in the top right corner indicates the orientation of the transmission axis of the linear polariser used during the experiment. The vertical and horizontal cross-sections of the experimental \textbf{k}-space emission pattern are plotted next to the CCD image, which correspond to directions parallel ($\parallel$) and perpendicular ($\perp$) to the polariser, respectively. Consequently, $\parallel$- and $\perp$-cuts in Fig. 1b can be referred to as \textit{p}- and \textit{s}-polarised emissions, respectively. Dashed lines indicate the selected regions of the corresponding  $\parallel$- and $\perp$-cuts.
	
	In order to determine the intrinsic emission dipole orientation of excitons in InSe (i.e. the IP-to-OP ratio) and to disentangle the effects that the dielectric multilayer environment has on the \textbf{k}-space emission intensity, we employ the analytical model proposed by Benisty \textit{et al}. \cite{benisty1998method} to model the \textbf{k}-dependent emission profile of bulk InSe. This model has been recently applied to simulate the back-focal plane emission for both the PL and the Raman emission of different 2D materials \cite{brotons2018engineering} (see Methods). Figure 1c shows normalised \textbf{k}-space emission patterns calculated for pure IP (left panel) and pure OP (right panel) distributions of dipoles, respectively, emitting at an energy of $\sim$1.244 eV (see Fig. 2a) and distributed all along the thickness of an 90 nm-thick InSe flake deposited on top of a SiO$_2$/Si substrate with a SiO$_2$ thickness of 105.3 nm. In the calculations, the effect of a detection polariser oriented along the $k_y$ direction has been included in order to reproduce the experimental conditions in Fig. 1b. As observed in these figures, the calculated \textbf{k}-space patterns for pure IP and pure OP distribution of dipoles are very different from each other. First, an overall higher emission intensity is observed for the \textbf{k}-space pattern corresponding to a pure IP distribution of dipoles, which can be understood by the larger emission outcoupling of IP dipoles in comparison to OP ones. Second, the \textbf{k}-space emission pattern for a distribution of pure OP dipoles shows a dark axis along the direction perpendicular to the polariser, which can be understood by the fact that a pure OP dipole does not emit \textit{s}-polarised light. Consequently, only \textit{p}-polarised emission contains information about both IP and OP dipoles. The presence of such a dark axis in the \textbf{k}-space pattern after a detection polariser represents a characteristic signature of a pure OP emission dipole.
	
	Blue and red dashed lines in the left and bottom panels of Fig. \ref{figure1}b represent the results of a simultaneous fit of the cross-sections of the experimental \textbf{k}-space pattern to the analytical model described in Methods. The fits reveal a 97$\%$ OP and 3$\%$ IP intrinsic emission dipole distribution in bulk InSe, which demonstrates the nearly complete OP emission characteristics of luminescent excitons in bulk InSe.
	
		\begin{figure*}[t]
		\begin{center}
		\includegraphics[scale=0.6]{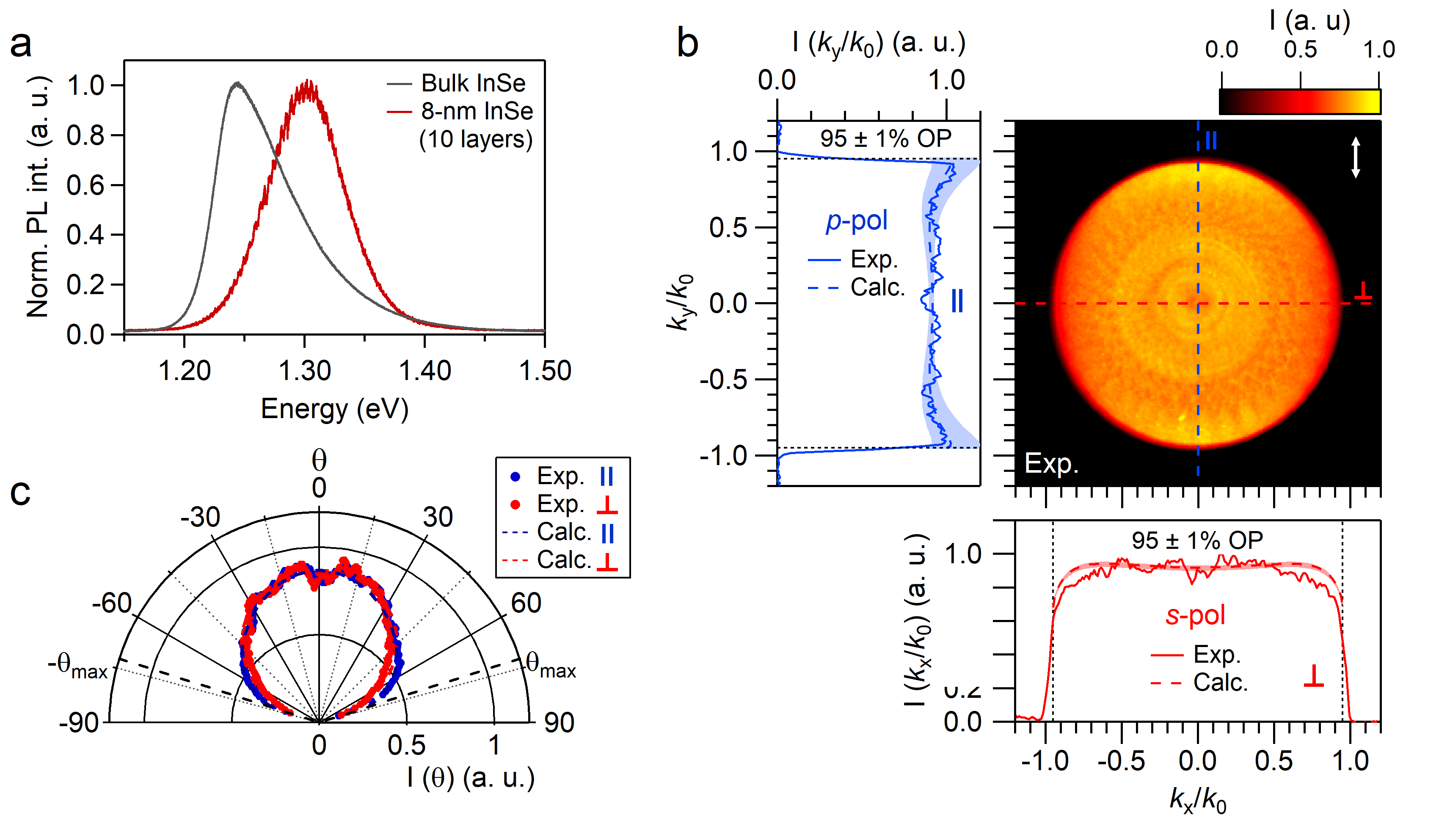}
		\end{center}
		\caption{\textbf{Orientation of luminescent excitons in 2D InSe.} (\textbf{a}) Comparison of the intensity-normalised room-temperature PL of bulk and 8-nm-thick InSe. (\textbf{b}) Intensity-normalised \textbf{k}-space emission pattern of a 8-nm-thick InSe flake deposited on top of a SiO$_2$/Si substrate with a SiO$_2$ thickness of 104.0 $\pm$ 0.1 nm as a function of the in-plane photon wavevector normalised to that in air. The orientation of the transmission axis of the linear polariser used during the experiment is indicated by the white arrow in the top right corner. Blue and red solid lines in the left and bottom panel show the experimental \textbf{k}-space cross-sections measured along the directions $\parallel$ and $\perp$ to the polariser (blue and red dashed lines in the CCD image, respectively). The blue and red dashed lines in the adjacent panels represent fittings of the experimental data to the analytical model described in Methods. The shadowed regions around the calculated values represent a confidence interval of 1$\%$ for the dipole orientation. (\textbf{c}) Measured and calculated far-field emission patterns for the studied InSe flake as a function of the emission angle $\theta$. The black dashed lines indicate the maximum collection angle ($\theta_{max}$) provided by the numerical aperture of the objective used in our experiments.}
		\label{figure2}
	\end{figure*}
	
	\subsection{Orientation of luminescent excitons in 2D InSe}
	In order to investigate whether 2D InSe retains the OP emission characteristic of the bulk material, we carried out measurements of the \textbf{k}-space emission pattern for an 8-nm-thick InSe flake. Figure 2a shows a comparison of the intensity-normalised room-temperature PL of bulk and 8-nm-thick InSe. A blueshift of $\sim$58 meV in the central emission energy is observed for the 2D InSe flake, a consequence of the quantum-confinement-induced increase of the electronic band gap of InSe with reducing thickness \cite{mudd2013tuning, sanchez2014electronic,brotons2016nanotexturing}, which confirms the 2D nature of the studied flake. Figure 2b shows the measured normalised \textbf{k}-space emission pattern of the 8-nm-thick InSe flake deposited on top of a SiO$_2$/Si substrate with a SiO$_2$ thickness of 104.0 $\pm$ 0.1 nm. The labelling in this figure is the same as the one employed in Fig. \ref{figure1}b. We observe that the \textit{s}- and \textit{p}-polarised \textbf{k}-space cross-sections for this sample (bottom and left panels of Fig. \ref{figure2}b, respectively) are much more similar to each other than for the bulk sample (see Fig. 1b). The origin of this behaviour lies in the angular redistribution of the emitted light induced by the different thicknesses and emission wavelengths of the bulk and 8-nm InSe flakes \cite{lukosz1979light}. A simultaneous fit of the measured \textbf{k}-space intensity profiles to the analytical model described in Methods reveals a 95 $\pm$ 1$\%$ OP intrinsic emission dipole distribution in the 8-nm-thick 2D InSe flake. The shadowed regions around the calculated values represent a confidence interval of 1$\%$ for the dipole orientation. These results demonstrate that 2D InSe retains the nearly complete OP dipole emission characteristics previously demonstrated for the bulk counterpart. Finally, the polar plot in Fig. 2c shows the measured and calculated far-field emission patterns for the studied InSe flake as a function of the emission angle $\theta$. The black dashed lines indicate the maximum collection angle ($\theta_{max}$) = 71.8$^\circ$ provided by the the objective with NA = 0.95 used in our experiments.

	\begin{figure*}[t]
		\begin{center}
		\includegraphics[scale=0.6]{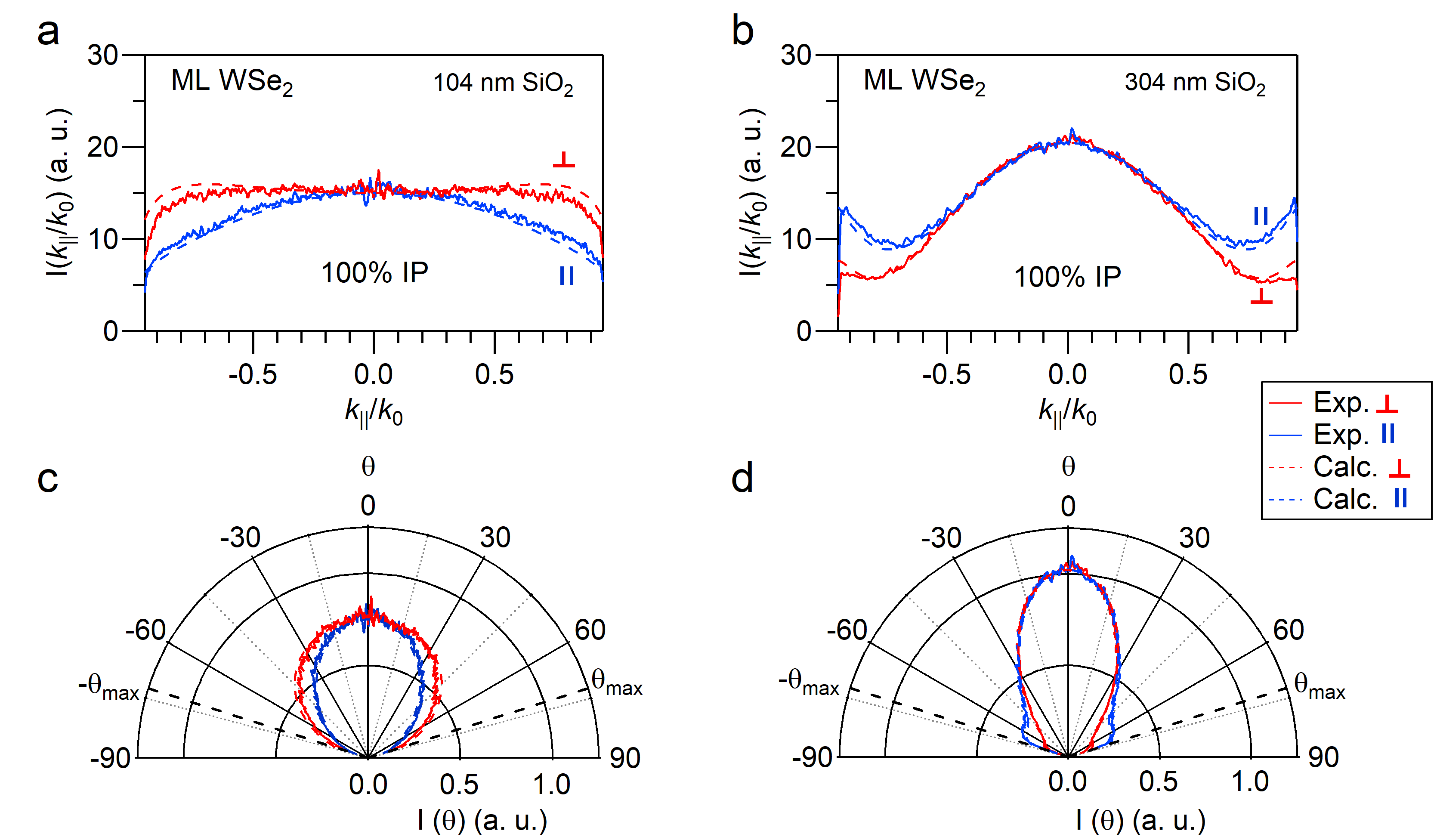}
		\end{center}
		\caption{\textbf{Orientation of luminescent excitons in ML WSe$_2$}. (\textbf{a})-(\textbf{b}) Experimental \textbf{k}-space emission profiles of ML WSe$_2$ flakes deposited on top of SiO$_2$/Si substrates with SiO$_2$ thicknesses of 104.0 $\pm$ 0.1 nm and 304.0 $\pm$ 0.1 nm, respectively, for directions $\parallel$ (blue solid lines) and $\perp$ (red solid lines) to the collection polariser. Dashed lines represent the result of a simultaneous fit of the experimental \textbf{k}-space profiles to the analytical model described in Methods. (\textbf{c})-(\textbf{d}) Measured and calculated far-field emission patterns corresponding to the results shown in Fig. \ref{figure3}a and Fig. \ref{figure3}b as a function of the emission angle $\theta$, respectively. The black dashed lines indicate the maximum collection angle ($\theta_{max}$) provided by the objective used in our experiments.}
		\label{figure3}
	\end{figure*}
	
	The orbital nature of the electronic states at the band-gap edges makes 2D InSe a valuable building-block in the design of van der Waals heterostructures with tailored optoelectronic properties. On the one hand, the strong $p_z$ orbital nature of the top valence band endows InSe with one of the largest band-gap tunability ranges found in a 2D semiconductor when its thickness varies between the bulk and the ML. On the other hand, electronic states of $p_z$ orbital nature are even under mirror symmetry operation; combining these states with the odd symmetry of the bottom of the conduction band determines the OP orientation of the luminescent excitons in this 2D semiconductor. 
	
	\subsection{Role of grey excitons in the emission of 2D MoSe$_2$ and WSe$_2$}
	Such an OP emission characteristic contrasts with the strongly IP-localised excitons of other 2D semiconductors such as 2D MoS$_2$ \cite{Schuller2013}, 2D black phosphorus \cite{Wang2015} and 2D CdSe nanoplatelets \cite{Scott2017, gao2017cdse, heckmann2017directed}. For 2D MoS$_2$, it has been demonstrated that the room-temperature PL originates solely from in-plane excitons \cite{Schuller2013}. Consequently, the PL emission of ML MoS$_2$ can be described by an isotropic distribution of incoherently radiating dipoles lying in planes parallel to the layer interfaces. The analogy of the nature of the orbitals involved in the optical band gaps of other Mo-based 2D TMDs such as MoSe$_2$ suggests analogously strong IP-localisation of luminescent excitons in this 2D van der Waals semiconductor at room temperature. Supplementary Figure 2 in the Supplementary Note 1 shows the experimental and calculated \textbf{k}-space emission pattern of ML MoSe$_2$ deposited on top of a gold substrate. Fitting of the experimental data to the analytical model reveals a pure IP dipole orientation also for 2D MoSe$_2$.
	
	Similar to 2D MoSe$_2$, it is usually assumed that the room-temperature PL emission of ML WSe$_2$ also originates from purely IP dipoles, due to the analogy of the electronic band structures of group-VI TMDs. However, Mo- and W-based TMDs present spin-orbit-split conductions bands with different sign at the $\pm K$ points of the Brillouin zone, which leads to an excitonic ground state with an IP transition dipole moment that is bright for Mo-based TMDs, and two energetically-split dark exciton states for W-based TMDs with OP transition dipole moments \cite{wang2018colloquium}. The upper dark energy state (grey exciton) is electric-dipole-allowed for electric fields polarized along the $z$-direction, while the low energy state is strictly dipole forbidden (perfectly dark) \cite{robert2017fine, zhang2017magnetic}. Low-temperature PL measurements of ML WSe$_2$ and WS$_2$ with high NA objectives have shown that grey excitons contribute significantly to the emitted PL signal \cite{wang2017plane}, making the low-temperature PL of these semiconductors originating not exclusively from IP dipoles but from a combination of IP and OP dipoles. However, a quantitative and unambiguous determination of the contribution of the grey excitons to the room-temperature PL of W-based TMDs is still missing.
	
	Therefore, we next experimentally quantify the contribution of OP dipoles to the room-temperature PL of ML WSe$_2$. Figures \ref{figure3}a and \ref{figure3}b show the experimental \textbf{k}-space emission profiles of ML WSe$_2$ flakes deposited on top of SiO$_2$/Si substrates with SiO$_2$ thicknesses of 104.0 $\pm$ 0.1 and 304.0 $\pm$ 0.1 nm, respectively, for directions $\parallel$ (blue solid line) and $\perp$ (red solid line) to the collection polariser. Due to interference effects, the \textbf{k}-space emission profiles are strongly affected by the SiO$_2$ thickness \cite{brotons2018engineering}. Dashed lines in Figs. \ref{figure3}a and \ref{figure3}b represent the result of a simultaneous fit of the experimental \textbf{k}-space profiles to the analytical model described in Methods. Figures 3c and 3d show the measured and calculated far-field emission patterns shown in Fig. 3a and Fig. 3b, respectively, as a function of the emission angle $\theta$. As can be seen in these figures, a very good agreement is observed between the experimental and the calculated values. The fits reveal a 100$\%$ IP intrinsic dipole distribution for ML WSe$_2$ deposited on both substrates. These results demonstrate the validity of the analytical model employed in this work to unravel the intrinsic emission dipole orientation of InSe. The good agreement between the experimental and calculated values observed in Fig. \ref{figure3} proves the capability of the employed analytical model to disentangle the dipole orientation from the angular redistribution of emitted light induced by the underlaying multilayer structure. More importantly, the results shown in Fig. \ref{figure3} indicate that the contribution of the grey exciton states (with OP dipoles) to the room-temperature PL of ML WSe$_2$ is negligible, since the far-field emission patterns can be reproduced with a distribution of pure IP dipoles. These results are supported by very recent theoretical calculations \cite{brem2019phonon}.
	
	\subsection{Electronic nature of the vertical dipole in InSe}
	
	In order to explain quantitatively the orientation of luminescent excitons in few-layer InSe we have performed \textit{ab initio} calculations using density functional theory within the local-density approximation as implemented in Quantum Espresso \cite{Giannozzi2009}. We have included the spin-orbit interaction with fully relativistic norm-conserving pseudopotentials \cite{epfl-pseudos} (see \footnote{The details of the converged parameters are: energy cutoff 90 Ry; k-point sampling $12 \times 12 \times 1$ in ML and $15 \times 15 \times10$ in bulk.}). In our calculations, we have assumed that the thickness of a ML InSe is, nominally, 0.833 nm \cite{rigoult1980refinement}.  
	
\begin{figure*}[t]
		\includegraphics[width=16 cm]{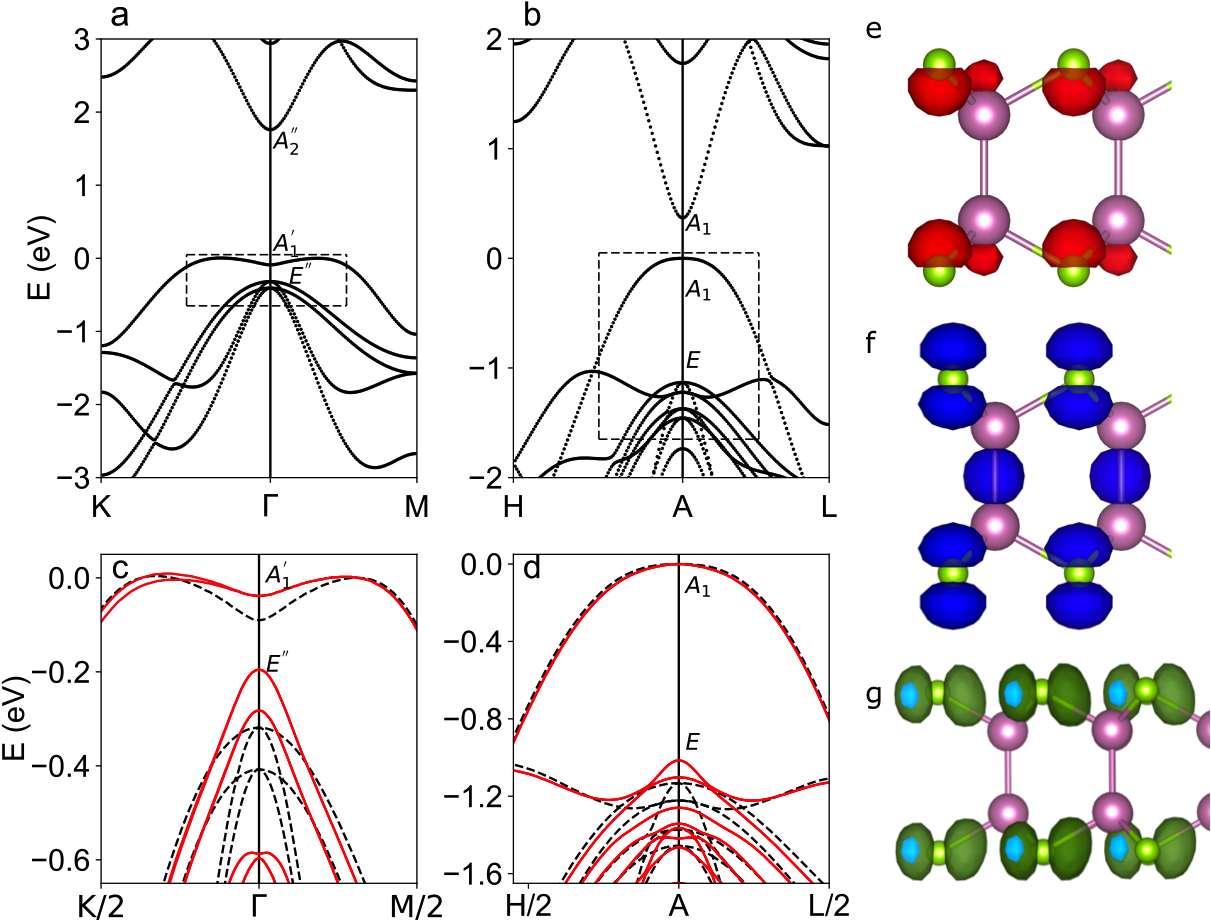}
		\caption{\textbf{Electronic properties of InSe}. Electronic band structure of (\textbf{a}) ML and (\textbf{b}) bulk InSealong the K-$\Gamma$-M and H-A-L high symmetry directions of the hexagonal Brillouin zone \cite{sanchez2014electronic}, respectively. Note that electronic states along K-$\Gamma$-M directions are equivalent to these along H-A-L  high symmetry directions in the case of ML InSe. The symmetry of initial and final states relevant for optical transitions are indicated in the plots. Details of the electronic structure at the top of the valence band for (\textbf{c}) ML and (\textbf{d}) bulk InSe, including the calculations without (black dashed lines) and with spin-orbit coupling (red solid lines). Wave functions at $\Gamma$ point for ML InSe corresponding to states at: (\textbf{e}) bottom of the conduction band ($A''_2$), (\textbf{f}) top of the valence band ($A'_1$), and (\textbf{g}) top of the second highest valence band ($E''$).}
		\label{theory-bands}
	\end{figure*}

	Figures \ref{theory-bands}a and \ref{theory-bands}b show the band structure of ML and bulk InSe, respectively, obtained without spin-orbit coupling. Regarding the ML InSe band structure, we can recognize the singular camel back around the $\Gamma$ point \cite{sanchez2014electronic}. We also indicate the simmetries of the bottom of the conduction band state ($A_2^{''}$) and the two upper valence band states ($A^{'}_1$ and $E^{''}$), which are key to identify allowed optical transitions. The selection rules for the dipole operator establish that, without spin-orbit splitting, transitions from the top of the valence band to the bottom of the conduction band are only allowed for $z$-axis (OP) polarised light. In the case of $(x,y)$-axis (IP) polarised light, only the transition from the $E^{''}$ to $A_2^{''}$ is allowed \cite{Magorrian2016,Zhou2017,brotons2016nanotexturing}. Table \ref{table1} shows the group representation of the valence and conduction band states at $\Gamma$. The even (1) or odd  (-1) character is an intuitive way to understand the selection rules. Thus, the product with the dipole operator is 1 (-1) for bright (dark) transitions. In the case of bulk InSe, the stacking order changes the symmetry of the crystal but keeps the same selection rules \cite{segura2018layered}.
	
	\begin{table}[b]
		\begin{tabular}{c | c | c | c | c | c}
			\hline
			\hline
			& $v_2$ & $v_1$ & $c_1$ & $\hat{x},\hat{y}$ & $\hat{z}$ \\
			\hline
			Bulk ($C_{3v}$)  & $E$ & $A_{1}$ & $A_{1}$ & $E$ & $A_{1}$ \\
			\hline
			ML ($D_{3h}$) & $E'' (-1)$ & $A_1' (1)$ & $A_2'' (-1)$ & $E' (1)$ & $A_2'' (-1)$\\
			\hline
			\hline
		\end{tabular}
		\label{table-symmetry}
		\caption{\textbf{Symmetry of states involved in optical transitions in InSe}. Group representation of the valence ($v_2$ and $v_1$) and conduction band ($c_1$) states for ML and bulk InSe, together with the representation of the dipole operators $\hat{x},\hat{y}$ and $\hat{z}$, corresponding to IP and OP emitting dipoles, respectively. The even and odd character of valence and conduction band states at $\Gamma$ point under in-plane mirror symmetry is indicated between brackets.}
		\label{table1}
	\end{table}
	
	Figures \ref{theory-bands}c and \ref{theory-bands}d show a detailed view of the valence band structure calculated with (solid lines) and without (dashed lines) spin-orbit interaction for ML and bulk InSe, respectively. The wave functions at $\Gamma$ of the conduction band state $A''_2 (c_1)$ and the valence band states $A'_1 (v_1)$ and $E'' (v_2)$ are plotted in Fig. \ref{theory-bands}e, f, and g for the ML case, which reflect their main In $s$, Se $p_z$ and Se $p_x-p_y$ orbital character, respectively. It is well-known that spin-orbit interaction makes IP optical transitions weakly allowed in bulk InSe by mixing with deeper lying valence bands \cite{brotons2016nanotexturing}. In fact, some mixing can be observed to occur (Fig. \ref{theory-bands}d) between the highest valence band (with $p_z$-character) and deeper valence bands (with $p_x-p_y$ character) when spin-orbit interaction is considered, although these bands are separated by $\sim$1 eV from each other. In ML InSe, band-structure calculations predict that these two set of valence bands are $\sim$0.3 eV apart from each other (Fig. \ref{theory-bands}c) and a stronger mixing of states between them is expected to occur. Consequently, our results reveal that the inclusion of spin-orbit interaction favours a mixing of valence band states with $p_z$ and $p_x-p_y$ orbital character that depends on the thickness of InSe, partly relaxing the restricted panorama of IP/OP allowed optical transitions in InSe described above (Table \ref{table1}).
	
	\begin{figure*}[t]
		\includegraphics[width=14 cm]{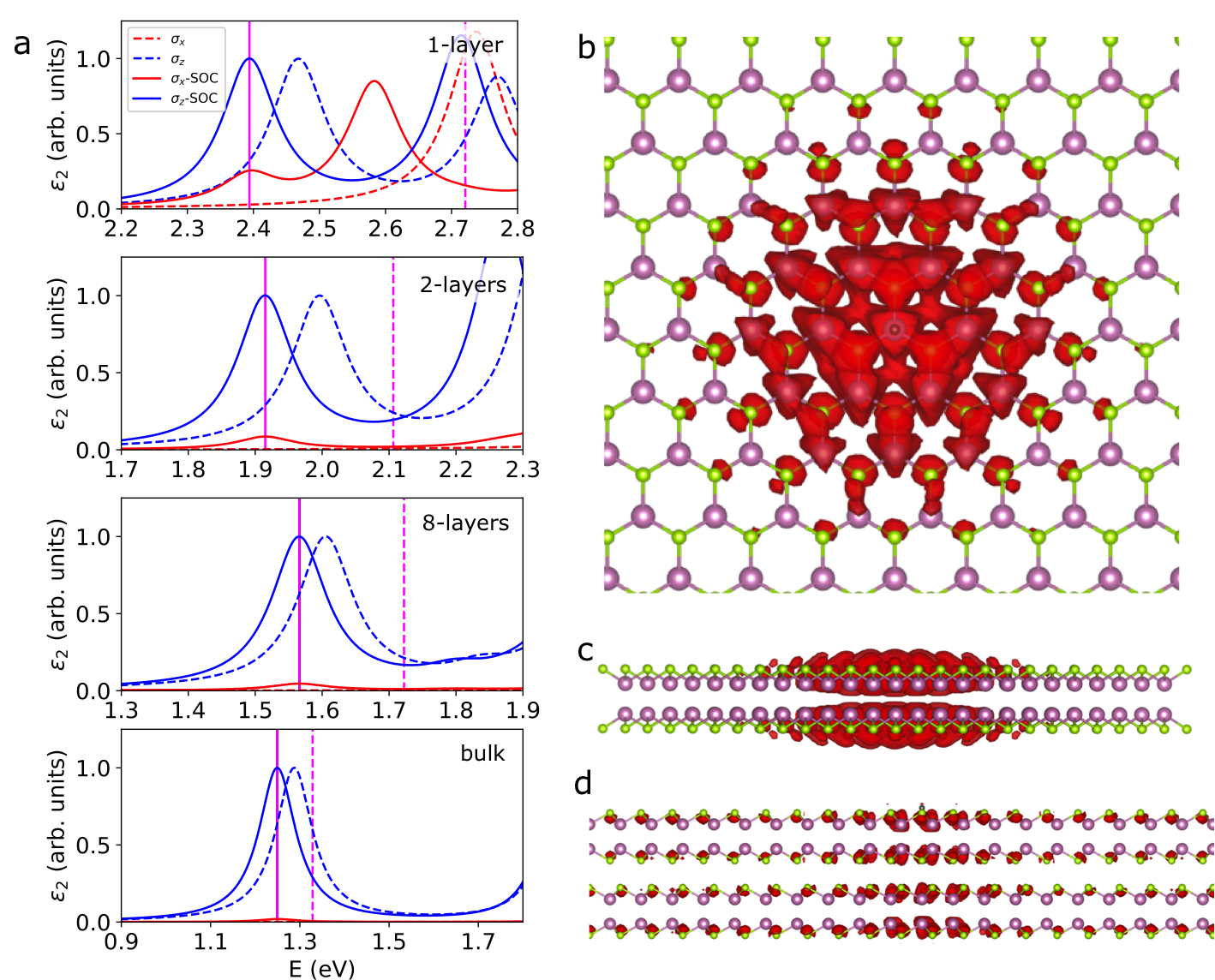}
		\caption{\textbf{Optical properties of 2D InSe}. (\textbf{a}) Optical absorption calculated in the independent-particle approximation for ML, bilayer, eight-layer, and bulk InSe for IP ($\sigma_x$) and OP ($\sigma_z$) polarised light. Solid (dashed) lines are calculations with (without) spin-orbit interaction. (\textbf{b}) Lateral and (\textbf{c}) surface density of the wave function of the first excitonic state of ML InSe. (\textbf{d}) Lateral density of the excitonic wave function of bilayer InSe.}
		\label{theory-ip}
	\end{figure*}
	
	In order to quantify the orientation of the luminescent excitons we have also computed the optical absorption including excitonic effects, within the framework of the Bethe-Salpeter Equation (BSE), as implemented in the Yambo code \cite{Marini2009, Sangalli2019}, with the aim to assure an accurate determination of the IP/OP character of excitons (see Supplementary Figures 4 and 5 and related discussion in Supplementary Note 3). At first approximation, we can consider the emission proportional to the absorption. Figure \ref{theory-ip}a shows the absorption of ML, bilayer, eight-layer, and bulk InSe, for IP ($\sigma_x$) and OP ($\sigma_z$) polarized light, with (solid lines) and without (dashed lines) spin-orbit interaction. Consistent with the previous discussion, optical absorption processes are either IP-allowed or OP-allowed in InSe without spin-orbit interaction. Spin-orbit interaction makes optical absorption processes involving states at the conduction and valence band edges slightly IP-allowed. Our calculations reveal that, when spin-orbit interactions are considered, a 2\% IP absorption can be expected at the band-gap energies (Fig. \ref{theory-ip}a), close to the value of 3\% obtained experimentally from our results in Fig. \ref{figure1}b. A similar situation is predicted to occur for InSe samples with important quantum-confinement effects. In particular, for 2D InSe of 8-layers thick, the IP contribution goes up to 4.7\%, which is also in very good agreement with the experimental increase of the IP contribution observed for thin InSe layers (5\% for 8-nm InSe -Fig. \ref{figure2}b-). In general terms, the IP absorption at band-gap energies becomes larger when the number of layers decreases \footnote{Band-gap is underestimated within the LDA approximation, and we have applied a scissor operator to match the experimental values. Nevertheless, the GW correction just produce a rigid-shift without implications in the spin-orbit interaction and the selection rules for IP and OP light\cite{antonius2018orbital}.}. Nevertheless, it is worth noting that for the absorption at the limit of a ML or bilayer, other effects like interaction with the substrate can also have a strong effect.
	
	We also show the excitonic wave function of the first excitonic state of ML InSe in Fig. \ref{theory-ip}b and \ref{theory-ip}c. We can appreciate the strong localisation in plane and the electronic density surrounding the selenium atoms. The consequence is a strong dependence on the number of layers due to the coupling of the wave functions of different layers \cite{sanchez2014electronic}. In the case of bilayer InSe, shown in Fig. \ref{theory-ip}d, the wave function spans over the different layers, supporting our hypothesis of the strong interlayer interaction. 
	
	It is worth mentioning that alternative calculation methods based on density functional theory (DFT) cannot determine the IP/OP exciton absorption ratio since DFT calculates optical spectra only within the independent-particle approximation excluding excitonic effects, although DFT can provide for an accurate estimate of the exciton binding energy. In contrast to this, the IP/OP absorption ratio, as estimated by BSE, is not changed due to excitonic effects. However, one of the drawbacks of the BSE is the convergence of the exciton binding energies with the number of \textbf{k}-points. While this convergence is not critical in 2D materials like ML or few-layer InSe, systems such as bulk InSe require a very large number of \textbf{k}-points. The convergence studies shown in Supplementary Figures 4 and 5 show the impact that a particular choice of the number of \textbf{k}-points has on the IP/OP absorption ratio and the binding energy for excitons in bulk InSe, respectively. These studies show that, although the exciton binding energy of bulk InSe converges very slowly with the number of \textbf{k}-points, the IP/OP exciton absorption ratio is already converged with the \textbf{k}-grid employed in our calculations (15x15x10), validating our results within the BSE framework.
	
	\section{Discussion}
	The strong IP-localisation of room-temperature luminescent excitons in some 2D semiconductors such as 2D MoS$_2$ \cite{Schuller2013}, MoSe$_2$, WSe$_2$, black phosphorus \cite{Wang2015}, or 2D CdSe nanoplatelets \cite{Scott2017, gao2017cdse, heckmann2017directed} is well suited for vertically emitting devices due to the large outcoupling efficiency of the radiation emitted by IP dipoles. In contrast, the IP orientation of the transition dipoles of these 2D semiconductors is not convenient for some planar photonics applications and for the study of light-matter interactions where dipole orientation is critical (such as dipole-dipole correlations). For some photonics applications it is instead desirable to have semiconductors with an OP orientation of the transition dipoles that benefit at the same time from the integration capabilities offered by the van der Waals platform. Examples of photonic applications in which OP dipoles outperform IP ones can be found for radiating dipoles embedded inside or placed at the surface of dielectric \cite{mulder2010dye, wang2016surface}, semiconductor \cite{jun2009broadband}, and metal-cladding optical waveguides \cite{rong2014spontaneous} with planar symmetry. In such planar photonics applications, OP dipoles not only present a higher coupling efficiency to the waveguide modes but also show a higher enhancement of the spontaneous emission rate (Purcell factor) than IP dipoles. Finally, the advantages of OP dipoles over IP ones for photonics applications are not restricted to planar waveguides. OP dipoles present higher Purcell enhancements and couple more efficiently than IP ones to multimode cylindrical waveguides and optical fibres when placed in the vicinity of the waveguides \cite{verhart2014single, davancco2009efficient, liebermeister2014tapered}.
	
	Our results show that the van der Waals semiconductor InSe fulfills the aforementioned properties. By performing \textbf{k}-space imaging of the far-field  photoluminescence signal of InSe flakes of different thicknesses, we demonstrate that not only bulk InSe but also 2D InSe flakes present dipole-allowed transition dipoles with a large intrinsic OP orientation. These findings are supported by \textit{ab-initio} calculations of the electronic band structure and the absorption coefficient of both bulk and 2D InSe. These results, together with the large band-gap tunability and the high electron mobility offered by 2D InSe, position the 2D forms of this semiconductor as valuable building-blocks in the design and fabrication of van der Waals heterostructures with tailored optoelectronic properties that can be harnessed in the next generation of photonics applications.
	
	\section{Methods}
	\subsection{Sample preparation}
	\begin{small}
		The InSe and WSe$_2$ flakes studied in this work were obtained by employing the mechanical exfoliation technique and then transferred on top of SiO$_{2}$/Si substrates. The thickness of the SiO$_2$ layers and the bulk InSe flake used in this work was determined by means of nulling ellipsometry measurements using a Nanofilm EP4 ellipsometer from Accurion. The ML thickness of the WSe$_2$ flake used in our experiments was confirmed by means of room-temperature photoluminescence measurements (see Supplementary Figure 3 in Supplementary Note 2). In a similar way, the thickness of the 2D InSe flake studied in this work was confirmed to be $\sim$ 8 nm by means of room-temperature PL, since it has been previously reported that the room-temperature PL emission energy of InSe can be used to efficiently determine the thickness of InSe flakes ranging between 1 and $\sim$14 nm \cite{mudd2013tuning,brotons2016nanotexturing}.
	\end{small}
	
	\subsection{\textbf{k}-space spectroscopy}
	\begin{small}
		\textbf{k}-space spectroscopy (or back-focal-plane imaging) was used to determine three-dimensional orientation of luminescent excitons in InSe. This technique allows to measure the far-field intensity distribution of PL, i.e. the PL signal as a function of the emission angle. For this, we used the back-focal plane imaging setup described in Supplementary Figure 1. A 532 nm continuous-wave excitation laser is collimated and reflected towards the sample with a $10\%$ reflective plate. A microscope objective with a NA = 0.95 then focuses the excitation laser on the sample in a diffraction limited spot. The PL emission from the material is then collected by the same microscope objective and collimated in the main path towards the collection optics and instruments. The two lenses in the main path focus the image of the back-focal plane of the microscope objective on a CCD camera, while keeping the beam collimated. A flip mirror in front of the CCD can be lifted to send the PL beam to a fibre and a spectrometer.
	\end{small}
	
	\begin{small}   
		To obtain a reliable result, the far-field measurement corresponding to the 2D material is performed behind a polariser which is rotated over $180^\circ$, with a \textbf{k}-space pattern acquired every $10^\circ$. The sample is then moved to image an area with bare substrate and the measurement is repeated without the 2D material to obtain a background measurement. After subtraction of the background, the $19$ images obtained are rotated and their average weighted by their integrated intensity is calculated. This process is used to mean out  any possibly existing anisotropy in the polarisation response of our setup, and also to mean out the imperfections and defaults present on the optics, to which the measurement is very sensitive.
	\end{small}
	
	\subsection{Model analysis of dipole emission}
	\begin{small}
		We calculate the \textbf{k}-dependent emission profile of the different 2D materials by employing the analytical model proposed by Benisty \textit{et al.} \cite{benisty1998method}, which has been recently applied to simulate the \textbf{k}-vector-dependent emission intensity for both the PL and the Raman emission of different 2D materials with very good results \cite{brotons2018engineering}. This model is based in the combination of a transfer matrix method and a dipole emission source term, and allows calculation of the emission of a thin source plane in an arbitrary planar structure in which the source plane is modelled by incoherent electrical oscillating dipoles. In this formalism, the \textbf{k}-space intensity profiles measured for directions perpendicular ($\perp$) and parallel ($\parallel$) to the collection polariser can be calculated as:
		\begin{align}
			\tilde{I}^{\perp}\left(k_{\parallel}\right)=C\alpha\tilde{I}^s_{\text{IP}}\left(k_{\parallel}\right)
			\label{Eq:k_space_s}
		\end{align}
		and
		\begin{align}
			\tilde{I}^{\parallel}\left(k_{\parallel}\right)=C[\alpha\tilde{I}^p_{\text{IP}}\left(k_{\parallel}\right)+\left(1-\alpha\right)\tilde{I}^p_{\text{OP}}\left(k_{\parallel}\right)],
			\label{Eq:k_space_p}
		\end{align}
		respectively. In these equations, $C$ is a proportionality constant that depends on different experimental parameters (such as integration time and excitation intensity), $\alpha$ represents the ratio between IP and OP emission dipoles, $k_{\parallel}$ is the in-plane photon momentum ($k_{\parallel}=k_0\sin(\theta)$) and the $s$ and $p$ superscripts denote TE and TM polarisations, respectively. Terms $\tilde{I}^{s,p}_{\text{IP}}\left(k_{\parallel}\right)$ and $\tilde{I}^{p}_{\text{OP}}\left(k_{\parallel}\right)$ in Equation (\ref{Eq:k_space_s}) and (\ref{Eq:k_space_p}) represent the \textbf{k}-space intensities calculated for distributions of purely IP and purely OP dipoles, respectively, in the corresponding experimental configuration. In our calculations, the effect of the excitation electric field has also been considered by following the formalism described in Refs. \cite{Aad2010enhancement, brotons2018engineering}. This formalism assumes the conservation of energy, and consequently, that the emission intensity of the dipole is proportional to the power dissipated by the corresponding dipole at the excitation energy. In the calculation of $\tilde{I}^{s,p}_{\text{IP}}\left(k_{\parallel}\right)$ and $\tilde{I}^{p}_{\text{OP}}\left(k_{\parallel}\right)$ we have also taken into account the influence of the thickness of the active material (important when dealing with thick flakes such as the bulk InSe flake used in our experiments). This was done by dividing the thickness of the active material into thin layers (sheets) of dipoles with thickness $\Delta z$ (with $\Delta z$ being the thickness of a ML for the corresponding material) and summing the emitted power from discrete subsources with the corresponding weighting coefficients.
		
		\par At this point, it is worth noticing that a close examination of Equation (\ref{Eq:k_space_s}) and Equation (\ref{Eq:k_space_p}) reveals that only the  \textbf{k}-space intensity profiles measured for directions parallel to the detection polariser contain information about the OP component of the intrinsic dipole orientation.
		
		\par Finally, in order to consider the broad PL emission of the materials studied in this work (in contrast to a purely monochromatic or narrow-band emission case), monochromatic \textbf{k}-space intensity calculations were preformed for emission wavelengths covering all the spectral range observed in the measured PL spectra and were then combined as follows:
		\begin{align}
			\tilde{I}^{s,p}_{\text{IP,OP}}\left(k_{\parallel}\right)=\int_{\lambda_{min}}^{\lambda_{max}}C_{\text{PL}}\left(\lambda\right)I^{s,p}_{\text{IP,OP}}\left(k_{\parallel},\lambda\right)\diff\lambda,
			\label{}
		\end{align}
		where $I^{s,p}_{\text{IP,OP}}\left(k_{\parallel},\lambda\right)$ is the \textbf{k}-space intensity calculated for a monochromatic emission wavelength $\lambda$ and $C_{\text{PL}}\left(\lambda\right)$ is a normalisation constant calculated from the experimental PL spectrum ($I_{\text{PL}}\left(\lambda\right)$) as
		\begin{align}
			C_{\text{PL}}\left(\lambda\right)=\frac{I_{\text{PL}}\left(\lambda\right)}{\int_{\lambda_{min}}^{\lambda_{max}}I_{\text{PL}}\left(\lambda\right)\diff\lambda}.
		\end{align}
		%.
	\end{small}
	
	\section{Data availability}
	Data described in this paper and presented in the Supplementary materials are available online at https://researchportal.hw.ac.uk/en/persons/brian-d-gerardot/datasets/
	
	\bibliographystyle{naturemag_noURL}
	\bibliography{Brotons-Gisbert_InSe_dipole_v2}

\begin{thebibliography}{10}
\expandafter\ifx\csname url\endcsname\relax
  \def\url#1{\texttt{#1}}\fi
\expandafter\ifx\csname urlprefix\endcsname\relax\def\urlprefix{URL }\fi
\providecommand{\bibinfo}[2]{#2}
\providecommand{\eprint}[2][]{\url{#2}}

\bibitem{Frenkel1931a}
\bibinfo{author}{Frenkel, J.}
\newblock \bibinfo{title}{{On the transformation of light into heat in solids.
  I}}.
\newblock \emph{\bibinfo{journal}{Phys. Rev.}} \textbf{\bibinfo{volume}{37}},
  \bibinfo{pages}{17} (\bibinfo{year}{1931}).

\bibitem{Yu2005Fundamentals}
\bibinfo{author}{Yu, P.~Y.} \& \bibinfo{author}{Cardona, M.}
\newblock \emph{\bibinfo{title}{{Fundamentals of Semiconductors: Physics and
  Materials Properties}}} (\bibinfo{publisher}{Springer},
  \bibinfo{year}{2005}).

\bibitem{chen2018theory}
\bibinfo{author}{Chen, H.-Y.}, \bibinfo{author}{Palummo, M.},
  \bibinfo{author}{Sangalli, D.} \& \bibinfo{author}{Bernardi, M.}
\newblock \bibinfo{title}{Theory and ab initio computation of the anisotropic
  light emission in monolayer transition metal dichalcogenides}.
\newblock \emph{\bibinfo{journal}{Nano Lett.}} \textbf{\bibinfo{volume}{18}},
  \bibinfo{pages}{3839--3843} (\bibinfo{year}{2018}).

\bibitem{Scott2017}
\bibinfo{author}{Scott, R.} \emph{et~al.}
\newblock \bibinfo{title}{{Directed emission of CdSe nanoplatelets originating
  from strongly anisotropic 2D electronic structure}}.
\newblock \emph{\bibinfo{journal}{Nat. Nanotechnol.}}
  \textbf{\bibinfo{volume}{12}}, \bibinfo{pages}{1155} (\bibinfo{year}{2017}).

\bibitem{wang2016surface}
\bibinfo{author}{Wang, Z.}, \bibinfo{author}{Zervas, M.~N.},
  \bibinfo{author}{Bartlett, P.~N.} \& \bibinfo{author}{Wilkinson, J.~S.}
\newblock \bibinfo{title}{{Surface and waveguide collection of Raman emission
  in waveguide-enhanced Raman spectroscopy }}.
\newblock \emph{\bibinfo{journal}{Opt. Lett.}} \textbf{\bibinfo{volume}{41}},
  \bibinfo{pages}{4146--4149} (\bibinfo{year}{2016}).

\bibitem{jun2009broadband}
\bibinfo{author}{Jun, Y.~C.}, \bibinfo{author}{Briggs, R.~M.},
  \bibinfo{author}{Atwater, H.~A.} \& \bibinfo{author}{Brongersma, M.~L.}
\newblock \bibinfo{title}{Broadband enhancement of light emission in silicon
  slot waveguides}.
\newblock \emph{\bibinfo{journal}{Opt. Exp.}} \textbf{\bibinfo{volume}{17}},
  \bibinfo{pages}{7479--7490} (\bibinfo{year}{2009}).

\bibitem{rong2014spontaneous}
\bibinfo{author}{Rong, T.} \& \bibinfo{author}{He, H.}
\newblock \bibinfo{title}{Spontaneous-emission coupling from an excited atom
  into a symmetrical metal-cladding optical waveguide}.
\newblock \emph{\bibinfo{journal}{Chin. Phys. Lett.}}
  \textbf{\bibinfo{volume}{31}}, \bibinfo{pages}{084205}
  (\bibinfo{year}{2014}).

\bibitem{verhart2014single}
\bibinfo{author}{Verhart, N.}, \bibinfo{author}{Lepert, G.},
  \bibinfo{author}{Billing, A.}, \bibinfo{author}{Hwang, J.} \&
  \bibinfo{author}{Hinds, E.}
\newblock \bibinfo{title}{Single dipole evanescently coupled to a multimode
  waveguide}.
\newblock \emph{\bibinfo{journal}{Opt. Exp.}} \textbf{\bibinfo{volume}{22}},
  \bibinfo{pages}{19633--19640} (\bibinfo{year}{2014}).

\bibitem{davancco2009efficient}
\bibinfo{author}{Davan{\c{c}}o, M.} \& \bibinfo{author}{Srinivasan, K.}
\newblock \bibinfo{title}{Efficient spectroscopy of single embedded emitters
  using optical fiber taper waveguides}.
\newblock \emph{\bibinfo{journal}{Opt. Exp.}} \textbf{\bibinfo{volume}{17}},
  \bibinfo{pages}{10542--10563} (\bibinfo{year}{2009}).

\bibitem{liebermeister2014tapered}
\bibinfo{author}{Liebermeister, L.} \emph{et~al.}
\newblock \bibinfo{title}{Tapered fiber coupling of single photons emitted by a
  deterministically positioned single nitrogen vacancy center}.
\newblock \emph{\bibinfo{journal}{Appl. Phys. Lett.}}
  \textbf{\bibinfo{volume}{104}}, \bibinfo{pages}{031101}
  (\bibinfo{year}{2014}).

\bibitem{molina2013effect}
\bibinfo{author}{Molina-S{\'a}nchez, A.}, \bibinfo{author}{Sangalli, D.},
  \bibinfo{author}{Hummer, K.}, \bibinfo{author}{Marini, A.} \&
  \bibinfo{author}{Wirtz, L.}
\newblock \bibinfo{title}{{Effect of spin-orbit interaction on the optical
  spectra of single-layer, double-layer, and bulk MoS$_2$}}.
\newblock \emph{\bibinfo{journal}{Phys. Rev. B}} \textbf{\bibinfo{volume}{88}},
  \bibinfo{pages}{045412} (\bibinfo{year}{2013}).

\bibitem{chernikov2014exciton}
\bibinfo{author}{Chernikov, A.} \emph{et~al.}
\newblock \bibinfo{title}{{Exciton binding energy and nonhydrogenic Rydberg
  series in monolayer WS$_2$}}.
\newblock \emph{\bibinfo{journal}{Phys. Rev. Lett.}}
  \textbf{\bibinfo{volume}{113}}, \bibinfo{pages}{076802}
  (\bibinfo{year}{2014}).

\bibitem{Schuller2013}
\bibinfo{author}{Schuller, J.~A.} \emph{et~al.}
\newblock \bibinfo{title}{Orientation of luminescent excitons in layered
  nanomaterials}.
\newblock \emph{\bibinfo{journal}{Nat. Nanotechnol.}}
  \textbf{\bibinfo{volume}{8}}, \bibinfo{pages}{271--276}
  (\bibinfo{year}{2013}).

\bibitem{wang2017plane}
\bibinfo{author}{Wang, G.} \emph{et~al.}
\newblock \bibinfo{title}{In-plane propagation of light in transition metal
  dichalcogenide monolayers: optical selection rules}.
\newblock \emph{\bibinfo{journal}{Phys. Rev. Lett.}}
  \textbf{\bibinfo{volume}{119}}, \bibinfo{pages}{047401}
  (\bibinfo{year}{2017}).

\bibitem{zhou2017probing}
\bibinfo{author}{Zhou, Y.} \emph{et~al.}
\newblock \bibinfo{title}{Probing dark excitons in atomically thin
  semiconductors via near-field coupling to surface plasmon polaritons}.
\newblock \emph{\bibinfo{journal}{Nat. Nanotechnol.}}
  \textbf{\bibinfo{volume}{12}}, \bibinfo{pages}{856} (\bibinfo{year}{2017}).

\bibitem{zhang2017magnetic}
\bibinfo{author}{Zhang, X.-X.} \emph{et~al.}
\newblock \bibinfo{title}{{Magnetic brightening and control of dark excitons in
  monolayer WSe$_2$}}.
\newblock \emph{\bibinfo{journal}{Nat. Nanotechnol.}}
  \textbf{\bibinfo{volume}{12}}, \bibinfo{pages}{883} (\bibinfo{year}{2017}).

\bibitem{Lu2017}
\bibinfo{author}{Lu, A.-Y.} \emph{et~al.}
\newblock \bibinfo{title}{Janus monolayers of transition metal
  dichalcogenides}.
\newblock \emph{\bibinfo{journal}{Nat. Nanotechnol.}}
  \textbf{\bibinfo{volume}{12}}, \bibinfo{pages}{744} (\bibinfo{year}{2017}).

\bibitem{Zhang2017}
\bibinfo{author}{Zhang, J.} \emph{et~al.}
\newblock \bibinfo{title}{{Janus Monolayer Transition-Metal Dichalcogenides}}.
\newblock \emph{\bibinfo{journal}{ACS Nano}} \textbf{\bibinfo{volume}{11}},
  \bibinfo{pages}{8192--8198} (\bibinfo{year}{2017}).

\bibitem{hong2014ultrafast}
\bibinfo{author}{Hong, X.} \emph{et~al.}
\newblock \bibinfo{title}{{Ultrafast charge transfer in atomically thin
  MoS$_2$/WS$_2$ heterostructures}}.
\newblock \emph{\bibinfo{journal}{Nat. Nanotechnol.}}
  \textbf{\bibinfo{volume}{9}}, \bibinfo{pages}{682} (\bibinfo{year}{2014}).

\bibitem{fang2014strong}
\bibinfo{author}{Fang, H.} \emph{et~al.}
\newblock \bibinfo{title}{Strong interlayer coupling in van der waals
  heterostructures built from single-layer chalcogenides}.
\newblock \emph{\bibinfo{journal}{Proc. Natl. Acad. Sci. U.S.A}}
  \textbf{\bibinfo{volume}{111}}, \bibinfo{pages}{6198--6202}
  (\bibinfo{year}{2014}).

\bibitem{Rivera2016}
\bibinfo{author}{Rivera, P.} \emph{et~al.}
\newblock \bibinfo{title}{{Valley-Polarized Exciton Dynamics in a 2D
  Semicondcutor Heterostructure}}.
\newblock \emph{\bibinfo{journal}{Science}} \textbf{\bibinfo{volume}{351}},
  \bibinfo{pages}{688--691} (\bibinfo{year}{2016}).

\bibitem{torun2018interlayer}
\bibinfo{author}{Torun, E.}, \bibinfo{author}{Miranda, H.~P.},
  \bibinfo{author}{Molina-S{\'a}nchez, A.} \& \bibinfo{author}{Wirtz, L.}
\newblock \bibinfo{title}{{Interlayer and intralayer excitons in MoS$_2$/WS$_2$
  and MoSe$_2$/WSe$_2$ heterobilayers}}.
\newblock \emph{\bibinfo{journal}{Phys. Rev. B}} \textbf{\bibinfo{volume}{97}},
  \bibinfo{pages}{245427} (\bibinfo{year}{2018}).

\bibitem{camassel1978excitonic}
\bibinfo{author}{Camassel, J.}, \bibinfo{author}{Merle, P.},
  \bibinfo{author}{Mathieu, H.} \& \bibinfo{author}{Chevy, A.}
\newblock \bibinfo{title}{Excitonic absorption edge of indium selenide}.
\newblock \emph{\bibinfo{journal}{Phys. Rev. B}} \textbf{\bibinfo{volume}{17}},
  \bibinfo{pages}{4718} (\bibinfo{year}{1978}).

\bibitem{mudd2013tuning}
\bibinfo{author}{Mudd, G.~W.} \emph{et~al.}
\newblock \bibinfo{title}{Tuning the bandgap of exfoliated {InSe} nanosheets by
  quantum confinement}.
\newblock \emph{\bibinfo{journal}{Adv. Mater.}} \textbf{\bibinfo{volume}{25}},
  \bibinfo{pages}{5714--5718} (\bibinfo{year}{2013}).

\bibitem{sanchez2014electronic}
\bibinfo{author}{S{\'a}nchez-Royo, J.~F.} \emph{et~al.}
\newblock \bibinfo{title}{Electronic structure, optical properties, and lattice
  dynamics in atomically thin indium selenide flakes}.
\newblock \emph{\bibinfo{journal}{Nano Res.}} \textbf{\bibinfo{volume}{7}},
  \bibinfo{pages}{1556--1568} (\bibinfo{year}{2014}).

\bibitem{brotons2016nanotexturing}
\bibinfo{author}{Brotons-Gisbert, M.} \emph{et~al.}
\newblock \bibinfo{title}{Nanotexturing to enhance photoluminescent response of
  atomically thin indium selenide with highly tunable band gap}.
\newblock \emph{\bibinfo{journal}{Nano Lett.}} \textbf{\bibinfo{volume}{16}},
  \bibinfo{pages}{3221--3229} (\bibinfo{year}{2016}).

\bibitem{andres2017quantum}
\bibinfo{author}{Andres-Penares, D.}, \bibinfo{author}{Cros, A.},
  \bibinfo{author}{Mart{\'\i}nez-Pastor, J.~P.} \&
  \bibinfo{author}{S{\'a}nchez-Royo, J.~F.}
\newblock \bibinfo{title}{Quantum size confinement in gallium selenide
  nanosheets: band gap tunability versus stability limitation}.
\newblock \emph{\bibinfo{journal}{Nanotechnology}}
  \textbf{\bibinfo{volume}{28}}, \bibinfo{pages}{175701}
  (\bibinfo{year}{2017}).

\bibitem{terry2018infrared}
\bibinfo{author}{Terry, D.~J.} \emph{et~al.}
\newblock \bibinfo{title}{{Infrared-to-violet tunable optical activity in
  atomic films of GaSe, InSe, and their heterostructures}}.
\newblock \emph{\bibinfo{journal}{2D Mater.}} \textbf{\bibinfo{volume}{5}},
  \bibinfo{pages}{041009} (\bibinfo{year}{2018}).

\bibitem{tamalampudi2014high}
\bibinfo{author}{Tamalampudi, S.~R.} \emph{et~al.}
\newblock \bibinfo{title}{{High performance and bendable few-layered InSe
  photodetectors with broad spectral response}}.
\newblock \emph{\bibinfo{journal}{Nano Lett.}} \textbf{\bibinfo{volume}{14}},
  \bibinfo{pages}{2800--2806} (\bibinfo{year}{2014}).

\bibitem{feng2015ultrahigh}
\bibinfo{author}{Feng, W.} \emph{et~al.}
\newblock \bibinfo{title}{{Ultrahigh photo-responsivity and detectivity in
  multilayer InSe nanosheets phototransistors with broadband response}}.
\newblock \emph{\bibinfo{journal}{J. Mater. Chem. C}}
  \textbf{\bibinfo{volume}{3}}, \bibinfo{pages}{7022--7028}
  (\bibinfo{year}{2015}).

\bibitem{mudd2015high}
\bibinfo{author}{Mudd, G.~W.} \emph{et~al.}
\newblock \bibinfo{title}{{High broad-band photoresponsivity of mechanically
  formed InSe--graphene van der Waals heterostructures}}.
\newblock \emph{\bibinfo{journal}{Adv. Mater.}} \textbf{\bibinfo{volume}{27}},
  \bibinfo{pages}{3760--3766} (\bibinfo{year}{2015}).

\bibitem{leisgang2018optical}
\bibinfo{author}{Leisgang, N.} \emph{et~al.}
\newblock \bibinfo{title}{{Optical second harmonic generation in encapsulated
  single-layer InSe}}.
\newblock \emph{\bibinfo{journal}{AIP Adv.}} \textbf{\bibinfo{volume}{8}},
  \bibinfo{pages}{105120} (\bibinfo{year}{2018}).

\bibitem{tonndorf2017single}
\bibinfo{author}{Tonndorf, P.} \emph{et~al.}
\newblock \bibinfo{title}{{Single-photon emitters in GaSe}}.
\newblock \emph{\bibinfo{journal}{2D Mater.}} \textbf{\bibinfo{volume}{4}},
  \bibinfo{pages}{021010} (\bibinfo{year}{2017}).

\bibitem{tonndorf2017chip}
\bibinfo{author}{Tonndorf, P.} \emph{et~al.}
\newblock \bibinfo{title}{On-chip waveguide coupling of a layered semiconductor
  single-photon source}.
\newblock \emph{\bibinfo{journal}{Nano Lett.}} \textbf{\bibinfo{volume}{17}},
  \bibinfo{pages}{5446--5451} (\bibinfo{year}{2017}).

\bibitem{mudd2016direct}
\bibinfo{author}{Mudd, G.} \emph{et~al.}
\newblock \bibinfo{title}{{The direct-to-indirect band gap crossover in
  two-dimensional van der Waals Indium Selenide crystals}}.
\newblock \emph{\bibinfo{journal}{Sci. Rep.}} \textbf{\bibinfo{volume}{6}},
  \bibinfo{pages}{39619} (\bibinfo{year}{2016}).

\bibitem{kuroda1980resonance}
\bibinfo{author}{Kuroda, N.} \& \bibinfo{author}{Nishina, Y.}
\newblock \bibinfo{title}{{Resonance Raman scattering study on exciton and
  polaron anisotropies in InSe}}.
\newblock \emph{\bibinfo{journal}{Solid State Commun.}}
  \textbf{\bibinfo{volume}{34}}, \bibinfo{pages}{481--484}
  (\bibinfo{year}{1980}).

\bibitem{kress1982cyclotron}
\bibinfo{author}{Kress-Rogers, E.}, \bibinfo{author}{Nicholas, R.},
  \bibinfo{author}{Portal, J.} \& \bibinfo{author}{Chevy, A.}
\newblock \bibinfo{title}{{Cyclotron resonance studies on bulk and
  two-dimensional conduction electrons in InSe}}.
\newblock \emph{\bibinfo{journal}{Solid State Commun.}}
  \textbf{\bibinfo{volume}{44}}, \bibinfo{pages}{379--383}
  (\bibinfo{year}{1982}).

\bibitem{segura1984electron}
\bibinfo{author}{Segura, A.}, \bibinfo{author}{Pomer, F.},
  \bibinfo{author}{Cantarero, A.}, \bibinfo{author}{Krause, W.} \&
  \bibinfo{author}{Chevy, A.}
\newblock \bibinfo{title}{Electron scattering mechanisms in n-type indium
  selenide}.
\newblock \emph{\bibinfo{journal}{Phys. Rev. B}} \textbf{\bibinfo{volume}{29}},
  \bibinfo{pages}{5708} (\bibinfo{year}{1984}).

\bibitem{bandurin2017high}
\bibinfo{author}{Bandurin, D.~A.} \emph{et~al.}
\newblock \bibinfo{title}{{High electron mobility, quantum Hall effect and
  anomalous optical response in atomically thin InSe}}.
\newblock \emph{\bibinfo{journal}{Nat. Nanotechnol.}}
  \textbf{\bibinfo{volume}{12}}, \bibinfo{pages}{223} (\bibinfo{year}{2017}).

\bibitem{Magorrian2016}
\bibinfo{author}{Magorrian, S.}, \bibinfo{author}{Z{\'o}lyomi, V.} \&
  \bibinfo{author}{Fal'ko, V.}
\newblock \bibinfo{title}{{Electronic and optical properties of two-dimensional
  InSe from a DFT-parametrized tight-binding model}}.
\newblock \emph{\bibinfo{journal}{Phys. Rev. B}} \textbf{\bibinfo{volume}{94}},
  \bibinfo{pages}{245431} (\bibinfo{year}{2016}).

\bibitem{li2018enhanced}
\bibinfo{author}{Li, Y.} \emph{et~al.}
\newblock \bibinfo{title}{Enhanced light emission from the ridge of
  two-dimensional inse flakes}.
\newblock \emph{\bibinfo{journal}{Nano Lett.}} \textbf{\bibinfo{volume}{18}},
  \bibinfo{pages}{5078--5084} (\bibinfo{year}{2018}).

\bibitem{song2018largely}
\bibinfo{author}{Song, C.} \emph{et~al.}
\newblock \bibinfo{title}{Largely tunable band structures of few-layer inse by
  uniaxial strain}.
\newblock \emph{\bibinfo{journal}{ACS Appl. Mater. Interfaces}}
  \textbf{\bibinfo{volume}{10}}, \bibinfo{pages}{3994--4000}
  (\bibinfo{year}{2018}).

\bibitem{lien2015engineering}
\bibinfo{author}{Lien, D.-H.} \emph{et~al.}
\newblock \bibinfo{title}{{Engineering light outcoupling in 2D materials}}.
\newblock \emph{\bibinfo{journal}{Nano Lett.}} \textbf{\bibinfo{volume}{15}},
  \bibinfo{pages}{1356--1361} (\bibinfo{year}{2015}).

\bibitem{brotons2018engineering}
\bibinfo{author}{Brotons-Gisbert, M.}, \bibinfo{author}{Mart{\'\i}nez-Pastor,
  J.~P.}, \bibinfo{author}{Ballesteros, G.~C.}, \bibinfo{author}{Gerardot,
  B.~D.} \& \bibinfo{author}{S{\'a}nchez-Royo, J.~F.}
\newblock \bibinfo{title}{Engineering light emission of two-dimensional
  materials in both the weak and strong coupling regimes}.
\newblock \emph{\bibinfo{journal}{Nanophotonics}} \textbf{\bibinfo{volume}{7}},
  \bibinfo{pages}{253--267} (\bibinfo{year}{2018}).

\bibitem{lieb2004single}
\bibinfo{author}{Lieb, M.~A.}, \bibinfo{author}{Zavislan, J.~M.} \&
  \bibinfo{author}{Novotny, L.}
\newblock \bibinfo{title}{Single-molecule orientations determined by direct
  emission pattern imaging}.
\newblock \emph{\bibinfo{journal}{J. Opt. Soc. Am. B}}
  \textbf{\bibinfo{volume}{21}}, \bibinfo{pages}{1210--1215}
  (\bibinfo{year}{2004}).

\bibitem{curto2010unidirectional}
\bibinfo{author}{Curto, A.~G.} \emph{et~al.}
\newblock \bibinfo{title}{Unidirectional emission of a quantum dot coupled to a
  nanoantenna}.
\newblock \emph{\bibinfo{journal}{Science}} \textbf{\bibinfo{volume}{329}},
  \bibinfo{pages}{930--933} (\bibinfo{year}{2010}).

\bibitem{lee2011planar}
\bibinfo{author}{Lee, K.} \emph{et~al.}
\newblock \bibinfo{title}{A planar dielectric antenna for directional
  single-photon emission and near-unity collection efficiency}.
\newblock \emph{\bibinfo{journal}{Nat. Photonics}}
  \textbf{\bibinfo{volume}{5}}, \bibinfo{pages}{166} (\bibinfo{year}{2011}).

\bibitem{bulgarini2014nanowire}
\bibinfo{author}{Bulgarini, G.} \emph{et~al.}
\newblock \bibinfo{title}{Nanowire waveguides launching single photons in a
  gaussian mode for ideal fiber coupling}.
\newblock \emph{\bibinfo{journal}{Nano Lett.}} \textbf{\bibinfo{volume}{14}},
  \bibinfo{pages}{4102--4106} (\bibinfo{year}{2014}).

\bibitem{gao2017cdse}
\bibinfo{author}{Gao, Y.}, \bibinfo{author}{Weidman, M.~C.} \&
  \bibinfo{author}{Tisdale, W.~A.}
\newblock \bibinfo{title}{{CdSe nanoplatelet films with controlled orientation
  of their transition dipole moment}}.
\newblock \emph{\bibinfo{journal}{Nano Lett.}} \textbf{\bibinfo{volume}{17}},
  \bibinfo{pages}{3837--3843} (\bibinfo{year}{2017}).

\bibitem{heckmann2017directed}
\bibinfo{author}{Heckmann, J.} \emph{et~al.}
\newblock \bibinfo{title}{Directed two-photon absorption in {CdSe}
  nanoplatelets revealed by k-space spectroscopy}.
\newblock \emph{\bibinfo{journal}{Nano Lett.}} \textbf{\bibinfo{volume}{17}},
  \bibinfo{pages}{6321--6329} (\bibinfo{year}{2017}).

\bibitem{benisty1998method}
\bibinfo{author}{Benisty, H.}, \bibinfo{author}{Stanley, R.} \&
  \bibinfo{author}{Mayer, M.}
\newblock \bibinfo{title}{Method of source terms for dipole emission
  modification in modes of arbitrary planar structures}.
\newblock \emph{\bibinfo{journal}{J. Opt. Soc. Am. A}}
  \textbf{\bibinfo{volume}{15}}, \bibinfo{pages}{1192--1201}
  (\bibinfo{year}{1998}).

\bibitem{lukosz1979light}
\bibinfo{author}{Lukosz, W.}
\newblock \bibinfo{title}{Light emission by magnetic and electric dipoles close
  to a plane dielectric interface. iii. radiation patterns of dipoles with
  arbitrary orientation}.
\newblock \emph{\bibinfo{journal}{J. Opt. Soc. Am.}}
  \textbf{\bibinfo{volume}{69}}, \bibinfo{pages}{1495--1503}
  (\bibinfo{year}{1979}).

\bibitem{Wang2015}
\bibinfo{author}{Wang, X.} \emph{et~al.}
\newblock \bibinfo{title}{Highly anisotropic and robust excitons in monolayer
  black phosphorus}.
\newblock \emph{\bibinfo{journal}{Nat. Nanotechnol.}}
  \textbf{\bibinfo{volume}{10}}, \bibinfo{pages}{517--521}
  (\bibinfo{year}{2015}).

\bibitem{wang2018colloquium}
\bibinfo{author}{Wang, G.} \emph{et~al.}
\newblock \bibinfo{title}{Colloquium: Excitons in atomically thin transition
  metal dichalcogenides}.
\newblock \emph{\bibinfo{journal}{Rev. Mod. Phys.}}
  \textbf{\bibinfo{volume}{90}}, \bibinfo{pages}{021001}
  (\bibinfo{year}{2018}).

\bibitem{robert2017fine}
\bibinfo{author}{Robert, C.} \emph{et~al.}
\newblock \bibinfo{title}{Fine structure and lifetime of dark excitons in
  transition metal dichalcogenide monolayers}.
\newblock \emph{\bibinfo{journal}{Phys. Rev. B}} \textbf{\bibinfo{volume}{96}},
  \bibinfo{pages}{155423} (\bibinfo{year}{2017}).

\bibitem{brem2019phonon}
\bibinfo{author}{Brem, S.} \emph{et~al.}
\newblock \bibinfo{title}{Phonon-assisted photoluminescence from dark excitons
  in monolayers of transition metal dichalcogenides}.
\newblock \emph{\bibinfo{journal}{Preprint at
  https://arxiv.org/abs/1904.04711}}  (\bibinfo{year}{2019}).

\bibitem{Giannozzi2009}
\bibinfo{author}{Giannozzi, P.} \emph{et~al.}
\newblock \bibinfo{title}{Quantum espresso: a modular and open-source software
  project for quantum simulations of materials}.
\newblock \emph{\bibinfo{journal}{J. Phys. Condens. Matter.}}
  \textbf{\bibinfo{volume}{21}}, \bibinfo{pages}{395502}
  (\bibinfo{year}{2009}).

\bibitem{epfl-pseudos}
\bibinfo{author}{Prandini, G.}, \bibinfo{author}{Marrazzo, A.},
  \bibinfo{author}{Castelli, I.~E.}, \bibinfo{author}{Mounet, N.} \&
  \bibinfo{author}{Marzari, N.}
\newblock \bibinfo{title}{Precision and efficiency in solid-state
  pseudopotential calculations}.
\newblock \emph{\bibinfo{journal}{npj Comput. Mater.}}
  \textbf{\bibinfo{volume}{7}} (\bibinfo{year}{2018}).

\bibitem{rigoult1980refinement}
\bibinfo{author}{Rigoult, J.}, \bibinfo{author}{Rimsky, A.} \&
  \bibinfo{author}{Kuhn, A.}
\newblock \bibinfo{title}{Refinement of the 3{R} $\gamma$-indium monoselenide
  structure type}.
\newblock \emph{\bibinfo{journal}{Acta Crystallogr. Sect. B-Struct. Sci.}}
  \textbf{\bibinfo{volume}{36}}, \bibinfo{pages}{916--918}
  (\bibinfo{year}{1980}).

\bibitem{Zhou2017}
\bibinfo{author}{Zhou, M.} \emph{et~al.}
\newblock \bibinfo{title}{Multiband
  $\mathbf{k}\ifmmode\cdot\else\textperiodcentered\fi{}\mathbf{p}$ theory of
  monolayer $x\mathrm{Se}$
  ($x=\mathrm{In},\phantom{\rule{0.28em}{0ex}}\mathrm{Ga}$)}.
\newblock \emph{\bibinfo{journal}{Phys. Rev. B}} \textbf{\bibinfo{volume}{96}},
  \bibinfo{pages}{155430} (\bibinfo{year}{2017}).

\bibitem{segura2018layered}
\bibinfo{author}{Segura, A.}
\newblock \bibinfo{title}{Layered indium selenide under high pressure: A
  review}.
\newblock \emph{\bibinfo{journal}{Crystals}} \textbf{\bibinfo{volume}{8}},
  \bibinfo{pages}{206} (\bibinfo{year}{2018}).

\bibitem{Marini2009}
\bibinfo{author}{Marini, A.}, \bibinfo{author}{Hogan, C.},
  \bibinfo{author}{Gr{\"u}ning, M.} \& \bibinfo{author}{Varsano, D.}
\newblock \bibinfo{title}{Yambo: an ab initio tool for excited state
  calculations}.
\newblock \emph{\bibinfo{journal}{Comput. Phys. Commun.}}
  \textbf{\bibinfo{volume}{180}}, \bibinfo{pages}{1392--1403}
  (\bibinfo{year}{2009}).

\bibitem{Sangalli2019}
\bibinfo{author}{Sangalli, D.} \emph{et~al.}
\newblock \bibinfo{title}{Many-body perturbation theory calculations using the
  yambo code}.
\newblock \emph{\bibinfo{journal}{J. Phys. Condens. Matter.}}
  \textbf{\bibinfo{volume}{31}}, \bibinfo{pages}{325902}
  (\bibinfo{year}{2019}).

\bibitem{mulder2010dye}
\bibinfo{author}{Mulder, C.~L.} \emph{et~al.}
\newblock \bibinfo{title}{Dye alignment in luminescent solar concentrators: I.
  vertical alignment for improved waveguide coupling}.
\newblock \emph{\bibinfo{journal}{Opt. Exp.}} \textbf{\bibinfo{volume}{18}},
  \bibinfo{pages}{A79--A90} (\bibinfo{year}{2010}).

\bibitem{Aad2010enhancement}
\bibinfo{author}{Aad, R.}, \bibinfo{author}{Blaize, S.},
  \bibinfo{author}{Bruyant, A.}, \bibinfo{author}{Couteau, C.} \&
  \bibinfo{author}{L{\'e}rondel, G.}
\newblock \bibinfo{title}{Enhancement of ultrathin film emission using a
  waveguiding active layer}.
\newblock \emph{\bibinfo{journal}{J. Appl. Phys.}}
  \textbf{\bibinfo{volume}{108}}, \bibinfo{pages}{123111}
  (\bibinfo{year}{2010}).

\bibitem{antonius2018orbital}
\bibinfo{author}{Antonius, G.}, \bibinfo{author}{Qiu, D.~Y.} \&
  \bibinfo{author}{Louie, S.~G.}
\newblock \bibinfo{title}{{Orbital Symmetry and the Optical Response of
  Single-Layer MX Monochalcogenides}}.
\newblock \emph{\bibinfo{journal}{Nano Lett.}} \textbf{\bibinfo{volume}{18}},
  \bibinfo{pages}{1925--1929} (\bibinfo{year}{2018}).

\end{thebibliography}
	
\section{Acknowledgements}

\begin{acknowledgments}
	
	This work was made possible by the Horizon 2020 research and innovation program (grant agreement No. 8204023). Work in Valencia was supported by the Spanish Government (Grant Nos. TEC2014-53727-C2-1-R) and the Comunidad Valenciana Government (Grant No. PROMETEOII/2014/059). Work in Edinburgh was supported by the EPSRC (EP/P029892/1) and the ERC (No. 725920). JFSR acknowledges financial support from the Salvador de Madariaga program (Grant PRX17/00186) of the Spanish Government. DAP acknowledges fellowship no. UV-INV-PREDOC17F1-539274 under the program "Atracci\'{o} de Talent, VLC-CAMPUS" of the University of Valencia. AMS acknowledges the Juan de la Cierva (Grant IJCI-2015-25799) program (MINECO, Spain). The computations were performed on the Tirant III cluster of the Servei d'Inform\`{a}tica of the University of Valencia (project vlc82). BDG thanks the Royal Society for a Wolfson Merit Award and the Royal Academy of Engineering for a Chair in Emerging Technology. The authors acknowledge Paul Dalgarno for providing technical assistance during the experimental measurements.
	
\end{acknowledgments}

\section{Author contributions}
M. B-G., R. Proux, B. D. G., and J. F. S-R. conceived the work. R. Picard, D. A-P., A. B., and J. F. S-R. fabricated the samples; R. Proux built the experimental setup and performed the \textbf{k}-space measurements with assistance from M. B-G.; M. B-G., R. Proux and R. Picard performed the ellipsometry characterisation of the samples; A. M-S. performed the \textit{ab initio} calculations of the electronic band structure and optical spectra; M. B-G. analysed the data with input from R. Proux; M. B-G., B. D. G., J. F. S-R., and A. M-S. wrote the manuscript with input from all authors; B. D. G. and J. F. S-R. supervised the project. M. B-G. and R. Proux contributed equally to this work.

\section{Competing Interests}
The authors declare no competing interests.

\end{document}